\documentclass[aps,prd,tightenlines,preprint,groupedaddress,showpacs,showkeys]{revtex4}

\usepackage{graphicx}

\begin{document}


\title{Quantum Gravity and Property of the very early Universe}

\author{Tomohiro Tsuneyama}
\email{tsune@hep.s.kanazawa-u.ac.jp}
\affiliation{Institute for Theoretical Physics,
 Kanazawa University, Kanazawa 920-1192, JAPAN}

\date{\today}

\begin{abstract}
As is well known, the universally accepted theory as quantum gravity 
(QG) doesn't exist. One of the main reasons for that is that quantized 
general relativity is perturbatively nonrenormalizable. 
But there are several theories whose low-energy effective action is 
general relativity such as super string theory, supergravity and so on. 
On the other hand there is the prospect that quantized general 
relativity will become consistent nonperturbatively, for example by way 
of the $\epsilon$ - expanded analysis in 2 + $\epsilon$ gravity 
and the implication of the exact renormalization group equation 
(ERGE) and so on. Thus, the reason that we can't check which is right, 
is because of no constraint from experiments. 
By the way, the early Universe can be considered a good laboratory for 
QG, because of the high energy scale. In this period it is thought that 
the effect of QG governed the Universe. Therefore to construct the 
Universe model out of consideration of a certain type of QG can be 
thought of the test of the theory. Here the result of quantized general 
relativity, which is treated nonperturbatively by ERGE, is checked. 
The application of QG improved Einstein equation to the very early 
Universe shows several characteristic properties. 
At first the Universe become free from the initial singularity. 
The second is the origin of the cosmological time is uniquely decided.
Furthermore this result doesn't destroy the success that the current 
cosmology attained.
\end{abstract}



\maketitle

\section{Introduction}
General relativity proposed by Einstein in 1915 can explain successfully 
the macroscopic properties of gravity. Nevertheless the effort trying 
to quantize general relativity failed because of the perturbative 
nonrenormalizability. Which is to say we can't control the 
divergence of perturbatively quantized general relativity.
As the result, general relativity is thought of 
as the only effective theory of a fundamental theory defined on 
higher energy scale. For example, super string theory and supergravity 
were proposed as such a theory. But it is shown that these theories are 
also perturbatively nonrenormalizable \cite{Deser}. 
On the other hand there is the prospect 
that quantized general relativity will become consistent 
nonperturbatively. By this means the $\epsilon$ - expanded analysis 
in 2 + $\epsilon$ gravity \cite{Kawai} and the implication of the 
ERGE \cite{Reuter} \cite{Souma} showed that quantized general relativity 
is nonperturbatively renormalizable according to the asymptotic 
safety that was put forward by Weinberg \cite{Weinberg}. 
This idea is that the quantum theory which lies on the ultraviolet 
critical surface of some ultraviolet fixed point is renormalizable. 
Thus several theories of QG exist, but an universally accepted one is 
absent. the reason of persisting in this situation is mainly because 
the constraints from experiments don't exist.

Currently, the measurements by cosmic microwave background (CMB) 
spectral, Type Ia supernovae observation and so on are consistent 
with an infinite flat universe containing about 30\% cold dark 
matter, 65\% dark energy and 5\% ordinary matter. Furthermore 
many observations are going to clarify the current and past Universe. 
In the circumstances the early Universe can be considered a good 
laboratory for QG. One of the reasons is to think that the Universe 
experienced the period such as immediately after the big bang 
in which QG effects dominated. And the trace of QG effects may be 
observed by the current precision measurements. Therefore to construct 
the Universe model out of consideration of a certain type of QG 
can be thought of as the test of the theory. If the model can 
explain some of the problems of modern cosmology such as the horizon 
and flatness problems, the cosmological constant problem, the one of 
the construction of the large scale structure, the identity of the 
dark matter and energy and so on, it is thought 
of as the suggestive matter to believe that the theory is correct.

Here the very early Universe is considered in according to the result 
by the analysis of the ERGE of quantized general relativity. 
This idea is also an approximate one to justify the present 
situation that the classical general relativity is used to the study 
of the early Universe and inflation mechanism and so on, 
because the analysis of the ERGE assure that general relativity is the 
fundamental theory.

In the next section the ERGE formalism for QG by Reuter is reviewed 
in brief. In section 3 the properties of QG 
in light of the ERGE analysis are described. In section 4 the
applications of QG to the very early Universe and 
the behavior of the QG improved Universe are discussed.
Section 5 is devoted to the conclusion.

\section{Formulation of Exact Renormalization Group Equation for QG}
\subsection{Quantization of Gravity}
In this subsection the Lagrangian of gravity ${\cal L}_{\rm GR}$ which 
is the general functional of the metric $\gamma_{\mu \nu}$ is 
considered. Then the classical action of gravity $S_{\rm GR}$ is 
invariant under the local general coordinate transformation. 
$S_{\rm GR}$ is quantized in a fixed Euclidean background field 
in $d$ dimensions. The metric $\gamma_{\mu \nu}$ is decomposed as 
$\gamma_{\mu\nu}=\bar{g}_{\mu\nu}+h_{\mu\nu}$. Here, $\bar{g}_{\mu\nu}$ 
denotes a fixed Euclidean background field and $h_{\mu\nu}$ denotes the 
fluctuations around the background. Now the general coordinate 
transformation is given by
\begin{eqnarray}
\delta h_{\mu\nu}&=&\pounds_v\gamma_{\mu\nu}
=v^{\rho}\partial_{\rho}\gamma_{\mu\nu}
+\partial_{\mu}v^{\rho}\cdot\gamma_{\rho\nu}
+\partial_{\nu}v^{\rho}\cdot\gamma_{\mu\rho}\nonumber\\
&\equiv&\nabla_{\mu}v_{\nu}+\nabla_{\nu}v_{\mu},\\
\delta\bar{g}_{\mu\nu}&=&0,
\end{eqnarray}
where $\pounds_v$ denotes the Lie derivative with respect to the vector
field $v^{\mu}$. Because the local general coordinate 
transformation is a kind of gauge transformation, to fix the gauge the 
BRST symmetry is introduced. The BRST transformation of the gravitation 
field $h_{\mu\nu}$ is given by
\begin{equation}
\mbox{\boldmath $\delta$}_{\bf B} h_{\mu\nu}=
\kappa^{-2}\pounds_C\gamma_{\mu\nu}.
\end{equation}
Here $C_{\mu\nu}$ denotes the ghost field and 
$\mbox{\boldmath $\delta$}_{\bf B}$ is an anticommuting operator of the 
BRST transformation. In addition $\kappa$ is a constant expressed in 
terms of the bare Newton constant $\bar{G}$ as 
$\kappa=(32\pi\bar{G})^{-1/2}$. The transformation of the other fields 
are given by
\begin{eqnarray}
\mbox{\boldmath $\delta$}_{\bf B}\bar{g}_{\mu\nu}&=&0,\nonumber\\
\mbox{\boldmath $\delta$}_{\bf B}C^{\mu}&=&
\kappa^{-2}C^{\nu}\partial_{\nu}C^{\mu},\nonumber\\
\mbox{\boldmath $\delta$}_{\bf B}\bar{C}_{\mu}&=&B_{\mu},\nonumber\\
\mbox{\boldmath $\delta$}_{\bf B}B_{\mu}&=&0,
\end{eqnarray}
where $\bar{C}_{\mu}$ and $B_{\mu}$ denote the anti-ghost field and the 
auxiliary field, respectively. Then the gauge fixing Lagrangian and the 
Faddeev - Popov ghost Lagrangian are introduced by 
\begin{eqnarray*}
{\cal L}_{\rm GF}&=&
\kappa B_{\mu}(F^{\mu}+\frac{\alpha\kappa}{2}B^{\mu}),\\
{\cal L}_{\rm GH}&=&
-\kappa\bar{C}_{\mu}(\mbox{\boldmath $\delta$} F^{\mu}).
\end{eqnarray*}
Here $F_{\mu}$ is defined to become linear with respect to 
$h_{\mu\nu}$ as
\begin{eqnarray*}
F_{\mu}=
\sqrt{2}\kappa{\cal F}^{\rho\sigma}_{\mu}[\bar{g}]h_{\rho\sigma},
\qquad{\cal F}^{\rho\sigma}_{\mu}=\delta^{\rho}_{\mu}\bar{D}^{\sigma}
-\frac{1}{2}\bar{g}^{\rho\sigma}\bar{D}_{\mu},
\end{eqnarray*}
then the gauges are fixed to the harmonic gauge.
Here $\bar{D}$ is the covariant derivative with respect to 
$\bar{g}_{\mu\nu}$ and $\alpha$ denotes the gauge parameter.

Now the generating functional for the connected Green's function is 
given by 
\begin{equation}
W[{\rm ES};\bar{g}]=
\ln\int{\cal D}\Phi\exp[-S[\Phi;\bar{g}]-S_{\rm ES}],
\end{equation}
where the subscript {\rm ES} denotes external sources and 
$\Phi=\{h,C,\bar{C}\}$ is the shorthand notation. 
Here the classical action $S[\Phi;\bar{g}]$ is given by
\begin{equation}
S[\Phi;\bar{g}]=S_{\rm GR}[h;\bar{g}]
+S_{\rm GF}[h;\bar{g}]+S_{\rm GH}[\Phi;\bar{g}],
\end{equation}
\begin{eqnarray}
S_{\rm GR}[h;\bar{g}]&=&\int d^dx{\cal L}_{\rm GR}[\bar{g}+h],\\
S_{\rm GF}[h;\bar{g}]&=&
\frac{1}{2\alpha}\int d^dx\sqrt{\bar{g}}
\bar{g}^{\mu\nu}F_{\mu}F_{\nu},\\
S_{\rm GH}[\Phi;\bar{g}]&=&
-\sqrt{2}\int d^dx\sqrt{\bar{g}}\bar{C}_{\mu}
(\delta^{\mu}_{\nu}\bar{D}^2+\bar{R}^{\mu}_{\ \nu})C^{\nu}.
\end{eqnarray}
In addition note that $B$ integral has still completed and 
$\bar{R}^{\mu}_{\ \nu}$ denotes Ricci tensor with respect to 
$\bar{g}_{\mu\nu}$.

\subsection{Average Action of Gravity}
First $t^{\mu\nu},\bar{\sigma}_{\mu},\sigma^{\mu}$ and
$\beta^{\mu\nu},\tau_{\mu}$ are introduced as the external sources that 
couple with $h_{\mu\nu},C^{\mu},\bar{C}_{\mu}$ and the BRST 
transformation of gravitation and ghost fields respectively. Then 
the scale dependent generating functional for the connected Green's 
function $W_k$ is expressed as
\begin{equation}
W_k[J;\beta,\tau;\bar{g}_{\mu\nu}]
=\ln\int{\cal D}\Phi\exp[
-S[\Phi;\bar{g}]-\Delta S_k[\Phi;\bar{g}]-S_{\rm ES}]
\end{equation}
in terms of the shorthand notation $J=\{t,\bar{\sigma},\sigma\}$. Here 
k is an arbitrary momentum infrared cutoff scale that satisfies 
$k\le\Lambda$, $\Lambda$ is a physical ultraviolet cutoff which is 
associated with the classical action of the theory.
In addition the external source action $S_{\rm ES}$ and the cutoff 
action $\Delta S_k[\Phi;\bar{g}]$ are given by
\begin{eqnarray}
S_{\rm ES}&=&
-\int d^dx\sqrt{\bar{g}}[
J\Phi+\beta^{\mu\nu}\pounds_C(\bar{g}_{\mu\nu}+h_{\mu\nu})
+\tau_{\mu}C^{\nu}\partial_{\nu}C^{\mu}],\\
\Delta S_k[\Phi;\bar{g}]&=&\frac{1}{2}\kappa^2\int d^dx\sqrt{\bar{g}}
h_{\mu\nu}(R_k^{\rm grav}[\bar{g}])^{\mu\nu\rho\sigma}h_{\rho\sigma}
+\sqrt{2}\int d^dx\sqrt{\bar{g}}
\bar{C}_{\mu}(R_k^{\rm gh}[\bar{g}])C^{\mu},
\end{eqnarray}
respectively. The cutoff function $R_k[\bar{g}]$ is expressed in terms 
of the dimensionless cutoff function $R^{(0)}(u)$ as
\begin{eqnarray*}
R_k^{grav}[\bar{g}]&=&Z_k^{\rm grav} k^2R^{(0)}(-\bar{D}^2/k^2),\\
R_k^{gh}[\bar{g}]&=&Z_k^{\rm gh} k^2R^{(0)}(-\bar{D}^2/k^2).
\end{eqnarray*}
$R^{(0)}(u)$ is arbitrary function which decides the 
property of the cutoff scheme and must interpolate smoothly between 
$R^{(0)}(0)=1$ and $R^{(0)}(u\rightarrow\infty)=0$. In addition 
$Z_k^{\rm grav}$ and $Z_k^{\rm gh}$ denote the renormalization factors 
of the gravitation field and the ghost field ,respectively. In particular 
$Z_k^{\rm grav}$ is the tensor with respect to the background metric and 
is given by $(Z_k^{\rm grav})^{\mu\nu\rho\sigma}=\bar{g}^{\mu\rho}
\bar{g}^{\nu\sigma}Z_k^{\rm grav}$ as the simplest case. Then classical 
fields $\bar{h}_{\mu\nu},\xi^{\mu},\bar{\xi}_{\mu}$ are given by 
\begin{equation}
\bar{h}_{\mu\nu}=
\frac{1}{\sqrt{\bar{g}}}\frac{\delta W_k}{\delta t^{\mu\nu}}
,\qquad
\xi^{\mu}=
\frac{1}{\sqrt{\bar{g}}}\frac{\delta W_k}{\delta\bar{\sigma}_{\mu}}
,\qquad
\bar{\xi}_{\mu}=
\frac{1}{\sqrt{\bar{g}}}\frac{\delta W_k}{\delta\sigma^{\mu}},
\end{equation}
respectively. Thus the average action of gravity is decided by using 
the modified Legendre transformation as 
\begin{eqnarray}
\Gamma_k[\bar{h},\xi,\bar{\xi};\beta,\tau;\bar{g}]&=&
-W_k[J;\beta,\tau;\bar{g}]
+\int d^dx\sqrt{\bar{g}}[t^{\mu\nu}\bar{h}_{\mu\nu}
+\bar{\sigma}_{\mu}\xi^{\mu}+\sigma^{\mu}\bar{\xi}_{\mu}]
-\Delta S_k[\bar{h},\xi,\bar{\xi};\bar{g}]\nonumber\\
&\equiv&
\widetilde{\Gamma}_k[\bar{h},\xi,\bar{\xi};\beta,\tau;\bar{g}]
-\Delta S_k[\bar{h},\xi,\bar{\xi};\bar{g}].\label{link1}
\end{eqnarray}
Then the classical metric $g_{\mu\nu}$ which is the counterpart of 
the quantum metric $\gamma_{\mu\nu}$ is defined as 
\begin{equation}
g_{\mu\nu}(x)\equiv\bar{g}_{\mu\nu}(x)+\bar{h}_{\mu\nu}(x).
\end{equation}
From now the average action is expressed in terms of $g_{\mu\nu}$ 
instead of $\bar{h}$ as
\begin{eqnarray}
\Gamma_k[\bar{h},\xi,\bar{\xi};\beta,\tau;\bar{g}]
&=&\Gamma_k[g-\bar{g},\xi,\bar{\xi};\beta,\tau;\bar{g}]\nonumber\\
&\equiv&\Gamma_k[g,\bar{g},\xi,\bar{\xi};\beta,\tau].
\end{eqnarray}

\subsection{Evolution Equation of Gravity}
Differentiating Eq.(\ref{link1}) with respect to $t=\ln(k/\Lambda)$, 
and after some calculations the evolution equation of gravity is given by
\begin{eqnarray}
\partial_t\Gamma_k&=&
\frac{1}{2}{\rm Tr}\left[\left(\Gamma_k^{(2)}
+\kappa^2R_k^{\rm grav}\right)_{\bar{h}\bar{h}}^{-1}
\left(\partial_t(\kappa^2R_k^{\rm grav})\right)_{\bar{h}\bar{h}}\right]
\nonumber\\
& &
-\frac{1}{2}{\rm Tr}\left[\left\{\left(\Gamma_k^{(2)}
+\sqrt{2}R_k^{\rm gh}\right)_{\bar{\xi}\xi}^{-1}
-\left(\Gamma_k^{(2)}+\sqrt{2}R_k^{\rm gh}\right)
_{\xi\bar{\xi}}^{-1}\right\}
\left(\partial_t(\sqrt{2}R_k^{\rm gh})\right)_{\bar{\xi}\xi}\right].
\label{link2}
\end{eqnarray}
Here ${\rm Tr}=\sum\int d^dx\sqrt{\bar{g}}\int d^dy\sqrt{\bar{g}}$ and 
$\sum$ denotes to figure out the sum with respect to the indices both of 
the gravitation and ghost fields.$\Gamma_k^{(2)}$ is the Hessian of 
$\Gamma_k$ with respect to the subscript. In addition this evolution 
equation must satisfy as the initial condition the average action at 
the physical UV cutoff scale $\Gamma_\Lambda$ which is given by
\begin{eqnarray}
\Gamma_\Lambda[g,\bar{g},\xi,\bar{\xi};\beta,\tau]&=&
S_{\rm GR}[g-\bar{g};\bar{g}]+S_{\rm GF}[g-\bar{g};\bar{g}]+
S_{\rm GH}[g-\bar{g},\xi,\bar{\xi};\bar{g}]\nonumber\\
& &
-\int d^dx\sqrt{\bar{g}}[
\beta^{\mu\nu}\pounds_{\xi}g_{\mu\nu}
+\tau_{\mu}\xi^{\nu}\partial_{\nu}\xi^{\mu}].\label{link3}
\end{eqnarray}

\subsection{Gauge Invariant theory space}
In general the ERGE formulation of the gauge theory can't conserve 
the gauge symmetry explicitly, because it involves the cutoff. 
So in an ordinary way, to make only the observable gauge invariant is 
as best we can. In other words the effective action 
$\Gamma\equiv\lim_{k\rightarrow 0}\Gamma_k$ must be brought to satisfy 
the Ward - Takahashi equation. Now this requirement leads the next 
condition.
\begin{eqnarray}
0&=&\langle\mbox{\boldmath $\delta$}_{\bf B}\Delta S_k
+\mbox{\boldmath $\delta$}_{\bf B} S_{\rm ES}\rangle\nonumber\\
&=&
\Bigl\langle\mbox{\boldmath $\delta$}_{\bf B}\Delta S_k
-\int d^dx\sqrt{\bar{g}}J
(\mbox{\boldmath $\delta$}_{\bf B}\Phi)\Bigr\rangle
\end{eqnarray}
Rewriting this condition in terms of the average action $\Gamma_k$ and 
after some calculations, so-called the modified Ward - Takahashi 
equation is obtained as 
\begin{equation}
\int d^dx\frac{1}{\sqrt{\bar{g}}}
\left[\frac{\delta\Gamma_k}{\delta\bar{h}_{\mu\nu}}
\frac{\delta\Gamma_k}{\delta\beta^{\mu\nu}}
+\frac{\delta\Gamma_k}{\delta\xi^{\mu}}
\frac{\delta\Gamma_k}{\delta\tau_{\mu}}\right]=Y_k.\label{link4}
\end{equation}
Here $Y_k$ become complicated expression as the follows 
\begin{eqnarray*}
Y_k&=&
\kappa^2{\rm Tr}\left[(R_k^{\rm grav})^{\mu\nu\rho\sigma}
(\Gamma_k^{(2)}+\kappa^2R_k^{\rm grav})
_{\bar{h}_{\rho\sigma}\varphi}^{-1}
\frac{\delta^2 \Gamma_k}{\sqrt{\bar{g}}\delta\varphi
\sqrt{\bar{g}}\delta\beta^{\mu\nu}}\right]\\
& &
-\sqrt{2}{\rm Tr}\left[R_k^{\rm gh}
(\Gamma_k^{(2)}+\sqrt{2}R_k^{\rm gh})_{\xi^{\mu}\varphi}^{-1}
\frac{\delta^2\Gamma_k}{\sqrt{\bar{g}}\delta\varphi
\sqrt{\bar{g}}\delta\tau_{\mu}}\right]\\
& &
+2\alpha^{-1}\kappa^2{\rm Tr}\left[
R_k^{\rm gh}{\cal F}^{\rho\sigma}_{\mu}
(\Gamma_k^{(2)}+\kappa^2R_k^{\rm grav})
_{\bar{h}_{\rho\sigma}\bar{\xi}_{\mu}}^{-1}\right],
\end{eqnarray*}
where $\varphi=\{\bar{h},\xi,\bar{\xi}]\}$ is a shorthand notation of 
the classical fields set. Then $Y_k$ goes to zero as $k\rightarrow 0$ 
because of $R_k^{\rm grav}=R_k^{\rm gh}=0$ in this limit. That is to say 
it insures that the effective action $\Gamma_{k\rightarrow 0}$ satisfy 
the usual Ward - Takahashi equation. By the way the exact solution of 
the evolution equation (\ref{link2}) satisfy the modified Ward - 
Takahashi equation (\ref{link4}), of course. But there is no guarantee 
that the approximate solutions of Eq.(\ref{link2}) do so. Therefore 
the solution space that satisfy the Eq.(\ref{link4}) is called the 
gauge invariant space and used as the indicator of approximate quality. 
Note that the gauge invariant space is different from the space which 
is spanned by all the gauge invariant operator.

\subsection{Approximation of Evolution Equation}
To solve the evolution equation(\ref{link2}) the infinite theory space 
must be truncated to the adequate subspace. As first approximation, 
the following subspace is considered.
\begin{enumerate}
\item The evolution of both the ghost action and external sources are 
      neglected. As a result, $S_{\rm GF}$ and $S_{\rm GH}$ in $\Gamma_k$ 
      are in the same form as that in the bare action.
\item $\Gamma_k$ is decomposed formally as
      \begin{eqnarray}
       \Gamma_k[g,\bar{g},\xi,\bar{\xi};\beta,\tau]&=&
       \bar{\Gamma}_k[g]+\hat{\Gamma}_k[g,\bar{g}]
       +S_{\rm GF}[g-\bar{g};\bar{g}]
       +S_{\rm GH}[g-\bar{g},\xi,\bar{\xi};\bar{g}]\nonumber\\
       & &
       -\int d^dx\sqrt{\bar{g}}[\beta^{\mu\nu}\pounds_{\xi}g_{\mu\nu}
       +\tau_{\mu}\xi^{\nu}\partial_{\nu}\xi^{\mu}],\label{link5}
      \end{eqnarray}
      where $\bar{\Gamma}_k[g]\equiv\Gamma_k[g,g,0,0;0,0]$ and 
      $\hat{\Gamma}_k$ contains the deviations for 
      $g_{\mu\nu}\neq\bar{g}_{\mu\nu}$ so that $\hat{\Gamma}_k[g,g]=0$. 
      $\hat{\Gamma}_k$ can be interpreted as the quantum corrections to 
      $S_{\rm GF}$, so $\hat{\Gamma}_k=0$ for all $k$ become a candidate 
      for the first approximation. Of course, this ansatz satisfies the 
      initial condition (\ref{link3}).
\end{enumerate}
In addition it is sure that this approximated subspace is the gauge 
invariant space, because of $\hat{\Gamma}_k[g,\bar{g}]=0$. Here $Y_k$ is 
given by
\begin{eqnarray*}
Y_k=
-\int d^dx\pounds_{\xi}g_{\mu\nu}
\frac{\delta\hat{\Gamma}_k[g,\bar{g}]}{\delta g_{\mu\nu}(x)}.
\end{eqnarray*}

Now inserting the approximated average action (\ref{link5}) into 
Eq.(\ref{link2}), the evolution equation in the gauge invariant theory 
space is given by
\begin{eqnarray}
\partial_t\Gamma_k[g,\bar{g}]&=&
\frac{1}{2}{\rm Tr}\left[(\kappa^{-2}\Gamma_k^{(2)}[g,\bar{g}]
+R_k^{\rm grav}[\bar{g}])^{-1}
(\partial_t R_k^{\rm grav}[\bar{g}])\right]\nonumber\\
& &
-{\rm Tr}\left[\left(-(\delta^{\mu}_{\nu}D^2+R^{\mu}_{\ \nu})
+R_k^{\rm gh}[\bar{g}]\right)^{-1}
(\partial_t R_k^{\rm gh}[\bar{g}])\right].\label{link6}
\end{eqnarray}
Here,
\begin{equation}
\Gamma_k[g,\bar{g}]\equiv\Gamma_k[g,\bar{g},0,0;0,0]
=\bar{\Gamma}_k[g]+S_{\rm GF}[g-\bar{g};\bar{g}],
\end{equation}
where $D$ and $R_{\mu\nu}$ are the covariant derivative and Ricci tensor
with respect to $g_{\mu\nu}$, respectively.

\subsection{Einstein - Hilbert Truncation}
Eq.(\ref{link6}) is still too difficult to solve actually. So to make 
the problem easier, the theory space is constrained moreover. At first, 
to take the classical action $S_{\rm GR}$ as the Einstein - Hilbert 
action will be the simplest approximation. Einstein - Hilbert action is 
given by
\begin{equation}
S_{\rm GR}=
\frac{1}{16\pi\bar{G}}\int d^dx\sqrt{g}\left[-R(g)+2\bar{\lambda}\right]
=2\kappa^2\int d^dx\sqrt{g}\left[-R(g)+2\bar{\lambda}\right].
\end{equation}
Here $\bar{G}$ and $\bar{\lambda}$ are the bare Newton constant and bare 
cosmological constant, respectively. These bare constants are rewrote to 
the scale dependent couplings in light of the field renormalization 
factor as
\begin{eqnarray*}
\bar{G}\rightarrow\bar{G}_k=\frac{1}{Z_{Nk}}\bar{G},\quad
\bar{\lambda}\rightarrow\bar{\lambda}_k.
\end{eqnarray*}
In addition to assign the scale dependence of gauge parameter $\alpha$ as
\begin{eqnarray*}
\alpha\rightarrow\alpha_k=\frac{1}{Z_{Nk}}\alpha.
\end{eqnarray*}
Then the approximated average action becomes 
\begin{eqnarray}
\Gamma_k[g,\bar{g}]&=&
2\kappa^2Z_{Nk}\int d^dx\sqrt{g}\Big[-R(g)+2\bar{\lambda}_k\Big]
\nonumber\\
& &
+\kappa^2Z_{Nk}\int d^dx\sqrt{\bar{g}}\bar{g}^{\mu\nu}
({\cal F}^{\alpha\beta}_{\mu}g_{\alpha\beta})
({\cal F}^{\rho\sigma}_{\nu}g_{\rho\sigma}).\label{link7}
\end{eqnarray}
Here note that $\kappa=(32\pi\bar{G})^{-1/2}$ is an independent 
constant of $k$ and gauge parameter is fixed to one. Then 
differentiating Eq.(\ref{link7}) with respect to $t$ and applying the 
condition $g_{\mu\nu}=\bar{g}_{\mu\nu}$ in order to delete the gauge 
fixing term. At last extracting the system of differential equations 
about $Z_{Nk}$ and $\bar{\lambda}_k$, the $\beta$ - functions of 
gravity are obtained.

\subsection{$\beta$-Functions of Gravity}
In according to the above discussions, in light of the dimensionless 
couplings of the Newton constant $g_k$ and the cosmological constant 
$\lambda_k$ defined as
\begin{eqnarray*}
g_k&=&k^{d-2}\bar{G}_k=k^{d-2}Z_{Nk}^{-1}\bar{G},\\
\lambda_k&=&k^{-2}\bar{\lambda}_k,
\end{eqnarray*}
the $\beta$ - functions of gravity are given by 
\begin{eqnarray}
\partial_t g_k&=&[d-2+\eta(k)]g_k\label{link8}\\
\partial_t\lambda_k&=&-[2-\eta(k)]\lambda_k\nonumber\\
&+&\frac{1}{2}g_k(4\pi)^{1-d/2}\left[2d(d+1)\Phi^1_{d/2}(-2\lambda_k)
-d(d+1)\eta(k)\tilde{\Phi}^1_{d/2}(-2\lambda_k)-8d\Phi^1_{d/2}(0)\right].
\label{link9}
\end{eqnarray}
Here and after the bars are omitted due not to confuse and the anomalous 
dimension $\eta(k)$ is given by 
\begin{equation}
\eta(k)=\frac{g_k B_1(d,\lambda_k)}{1-g_k B_2(d,\lambda_k)},
\end{equation}
\begin{eqnarray*}
B_1(d,\lambda_k)&=&
\frac{1}{3}(4\pi)^{1-d/2}\Big[d(d+1)\Phi_{d/2-1}^1(-2\lambda_k)
-6d(d-1)\Phi_{d/2}^2(-2\lambda_k)\\
& &-4d\Phi_{d/2-1}^1(0)-24\Phi_{d/2}^2(0)\Big],\\
B_2(d,\lambda_k)&=&-\frac{1}{6}(4\pi)^{1-d/2}
\left[d(d+1)\tilde{\Phi}_{d/2-1}^1(-2\lambda_k)
-6d(d-1)\tilde{\Phi}_{d/2}^2(-2\lambda_k)\right].
\end{eqnarray*}
Here $\Phi$s are defined as the integral functions of the arbitrary 
dimensionless cutoff function $R^{(0)}$ as
\begin{eqnarray*}
\Phi_n^p(w)&=&\frac{1}{\Gamma(n)}\int_0^{\infty}dz z^{n-1}
\frac{R^{(0)}(z)-zR^{(0)}{}'(z)}{[z+R^{(0)}(z)+w]^p},\\
\tilde{\Phi}_n^p(w)&=&\frac{1}{\Gamma(n)}\int_0^{\infty}dz z^{n-1}
\frac{R^{(0)}(z)}{[z+R^{(0)}(z)+w]^p},
\end{eqnarray*}
and satisfy
\begin{eqnarray*}
\Phi_n^p(w)=\tilde{\Phi}_n^p(w)=\frac{1}{(1+w)^p}\qquad
{\rm as}\qquad n\rightarrow 0.
\end{eqnarray*}

\section{Property of QG}
\begin{figure}
\includegraphics{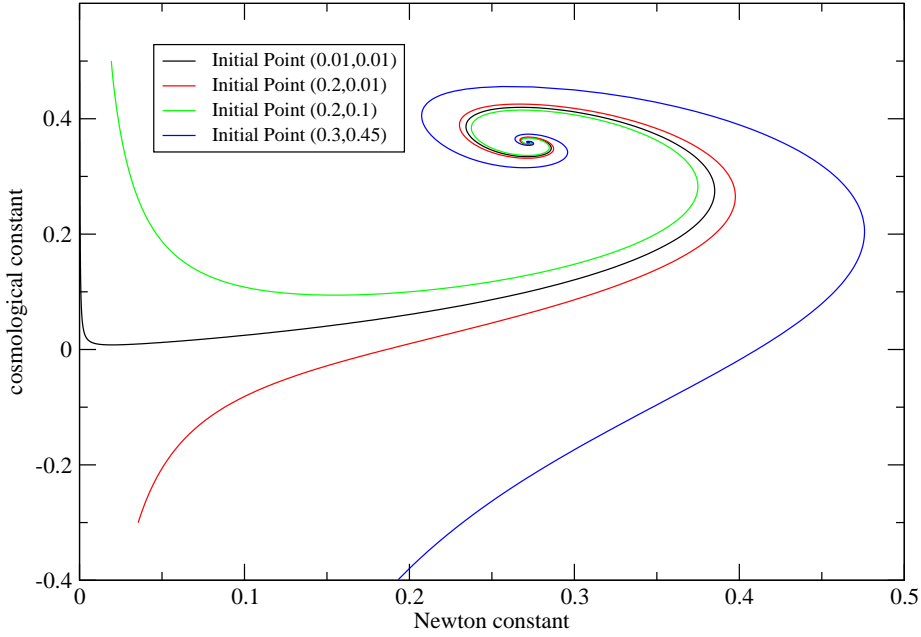}
\caption{The numerically calculated flows of the $\beta$ - functions for 
the Newton constant and the cosmological constant in four dimensions. 
The initial points in the figure denote the only starting points to draw 
the associated flows, so don't have the physical significance.}
\end{figure}
Fig.1 shows the numerically calculated flows of Eqs.(\ref{link8}) and 
(\ref{link9}) in four dimensions on the plane that is spanned by the 
$\sqrt{g}$ and $\sqrt{g} R$ operators. Explicitly two fixed point exist, 
these are an ultraviolet non - Gaussian fixed point (NGFP) and an infrared 
Gaussian fixed point (GFP). In particular the NGFP is attractive in the 
all directions as energy scale grows. This behavior is considered as the 
suggestion that $\sqrt{g}$ - $\sqrt{g} R$ plane is the subspace of
the ultraviolet critical surface. In this way according to the 
asymptotic safe argument, quantized general relativity in four dimension 
is expected to be renormalizable nonperturbatively. It is known that 
this structure is maintained even if the matter fields are included 
\cite{Percacci}, or $\sqrt{g} R^2$ term is taken into account.

In ultraviolet limit, in which is considered as the state of the very 
early Universe, the behaviors both of the Newton constant and the 
cosmological constant is uniquely controlled by the ultraviolet NGFP. 
That is to say that in neighborhood of the NGFP the dimensionful 
Newton constant $G_k$ and cosmological one $\lambda_k$ behave 
$G_k=g^*/k^\epsilon$ and $\lambda_k=\lambda^*k^\epsilon$ in 
$2+\epsilon$ dimensions, respectively. Here $g^*$ and $\lambda^*$ are 
the values of the dimensionless Newton constant and the cosmological 
one at the ultraviolet NGFP, respectively. Especially in four 
dimensions $G_k=g^*/k^2$ denotes that QG is asymptotic free.

\begin{figure}
\includegraphics{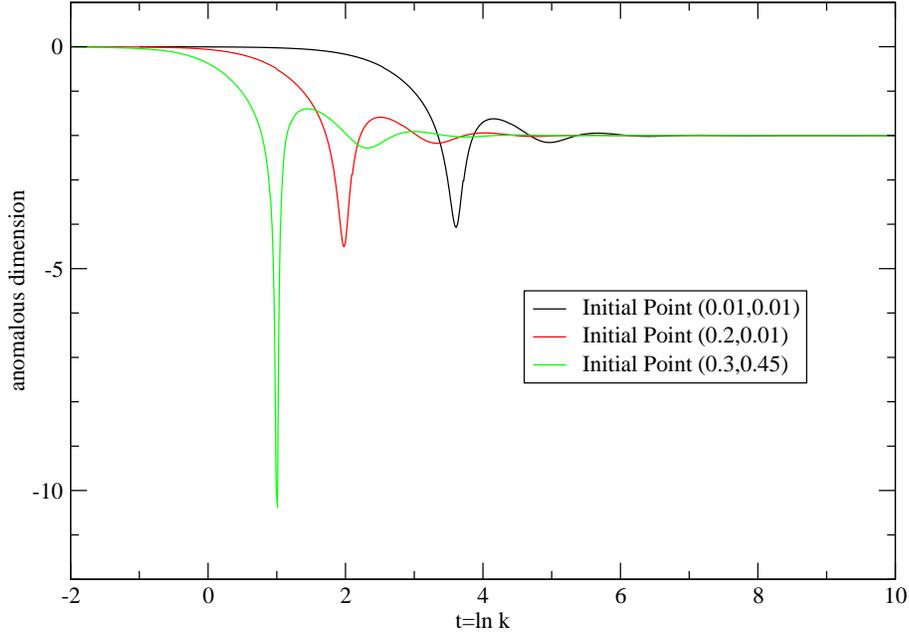}
\caption{The numerically calculated result of the anomalous dimension 
from the $\beta$ - functions in four dimensions. The initial points in 
the figure are the same as Fig.1.}
\end{figure}
On the other hand, in infrared region the dimensionful Newton constant 
$G_k$ behave the same as the classical general relativity exactly 
because of the anomalous dimension $\eta=0$ that is showed by Fig.2. 
But the same time the property of the dimensionful cosmological constant 
can't be controlled. It is thought as the disadvantage in light of the 
cosmological constant problem.

About the formalism used in this paper, the cutoff scheme dependence and 
the gauge parameter dependence are reported \cite{Reuter2} \cite{Souma2}. 
These may break the structure of the flow for the $\beta$ - functions. 
To solve these problems it is necessary the further investigation to 
include the other operators such as $\sqrt{g} R^2$ term and to consider 
the running gauge parameter in the approximate scheme and so on.

\section{Application to the very Early Universe}
In this section the result of the ERGE analysis for QG is applied the 
Robertson - Walker metric
\begin{equation}
ds^2=
-dt^2+a^2(t)\left[\frac{dr^2}{1-Kr^2}
+r^2(d\theta^2+\sin^2\theta d\phi^2)\right],
\end{equation}
which represents the homogeneous and isotropic Universe. Here $a(t)$ 
is the cosmological time $t$ dependent scale factor and $K$ is the 
curvature of the Universe. Though the time dependence of $a(t)$ follows 
the Einstein equation, to decide the dependence in the very early 
Universe the quantum effects of QG must be considered. The above ERGE 
analysis of QG indicates that the Newton constant $G$ and the 
cosmological constant $\Lambda$ in the classical Einstein equation must 
be treated as the scale dependent variables $G(k)$, $\Lambda(k)$. 
Therefore the ERGE modified Einstein equation is given by 
\begin{eqnarray}
G^{\mu}_{\ \nu}+\Lambda(k) \delta^{\mu}_{\ \nu}=
8\pi G(k)T^{\mu}_{\ \nu},\\
T^{\mu}_{\ \nu}\equiv(-\rho,p,p,p),
\end{eqnarray}
where $T^{\mu}_{\ \nu}$ represents the energy - momentum tensor of the 
Universe as the ideal fluid. Here the behaviors of $G(k)$ and 
$\Lambda(k)$ in the very early Universe is considered as
\begin{eqnarray}
G=g_*^{UV}/k^2,\\
\Lambda=\lambda_*^{UV}k^2,
\end{eqnarray}
in terms of the characteristic energy scale $k$ of a moment of the 
Universe. Secondly it is important how to identify the energy scale 
with the actual physical entity. Here I will introduce the energy scale 
related the actual geodesic distance. Namely, $k$ is expressed in terms 
of the comoving distance $x$ as 
\begin{equation}
k=\xi/xa.\label{link10}
\end{equation}
Here $\xi$ is a positive constant parameter. This identification has the 
advantage of that both the gravitational quantum effect observed in the 
today's Universe and in the very early Universe are treated in the same 
framework. For example if a cosmological characteristic scale such as 
the scale of the cluster of galaxies is considered, the quantum effect 
is dominant at when the scale factor is small. On the other hand 
concerning a fixed scale factor the quantum effect manifests at the 
very short length. In according to this assumption, the equations for 
the evolution of the very early Universe are given by
\begin{eqnarray}
\left(\frac{\dot{a}}{a}\right)^2=\frac{8\pi}{3}G(k)\rho
-\frac{K}{a^2}+\frac{\Lambda(k)}{3},\label{link11}\\
G(k)\left(-\dot{\rho}-3\frac{\dot{a}}{a}(\rho+p)\right)-\dot{G}(k)\rho
-\frac{\dot{\Lambda}(k)}{8\pi}=0,\label{link12}\\
G(k)=\widetilde{g_*}a^2,\qquad
\Lambda(k)=\widetilde{\lambda}_*/a^2.
\end{eqnarray}
Here $\widetilde{g_*}$ and $\widetilde{\lambda}_*$ are given by 
\begin{eqnarray*}
\widetilde{g_*}=g_*^{UV}x^2/\xi^2\qquad ,\qquad
\widetilde{\lambda}_*=\lambda_*^{UV}\xi^2/x^2,
\end{eqnarray*}
with respect to a fixed comoving distance, respectively. Note that 
the scale dependences of the Newton constant and the cosmological 
constant are given by not hand but the ERGE analysis.

These equations (\ref{link11}), (\ref{link12}) represent the very 
early Universe as followings.
\begin{enumerate}
\item $a$ - $\rho$ relation\\
The relation between the scale factor $a$ and the energy density $\rho$ 
of the Universe is derived from Eqs.(\ref{link11}) and (\ref{link12}). 
At first $K$, $\widetilde{g_*}$ and $\widetilde{\lambda}_*$ are defined 
as the appropriate constants. But five variables $(G,\Lambda,a,\rho,p)$ 
still remain whereas there are no more than four equations. Therefore the 
another equation must be given. Here though the relation is unknown, 
$\rho=wp$ is supposed as the equation of state. Then the relation 
between $a$ and $\rho$ is given by
\begin{equation}
\frac{d\rho}{\rho}=(-3w-5)\frac{da}{a}+\frac{1}{4\pi}
\frac{\widetilde{\lambda}_*}{\widetilde{g_*}}\frac{da}{a^5\rho}.
\label{link13}
\end{equation}
Secondly substituting $\rho$ with 
$\bar{\rho}=4\pi(3w+5)\widetilde{g_*}\rho/\widetilde{\lambda}_*$ 
Eq.(\ref{link13}) becomes
\begin{equation}
\frac{d\bar{\rho}}{da}=
(3w+5)\left(-\frac{\bar{\rho}}{a}+\frac{1}{a^5}\right).
\end{equation}
\begin{figure}
\includegraphics{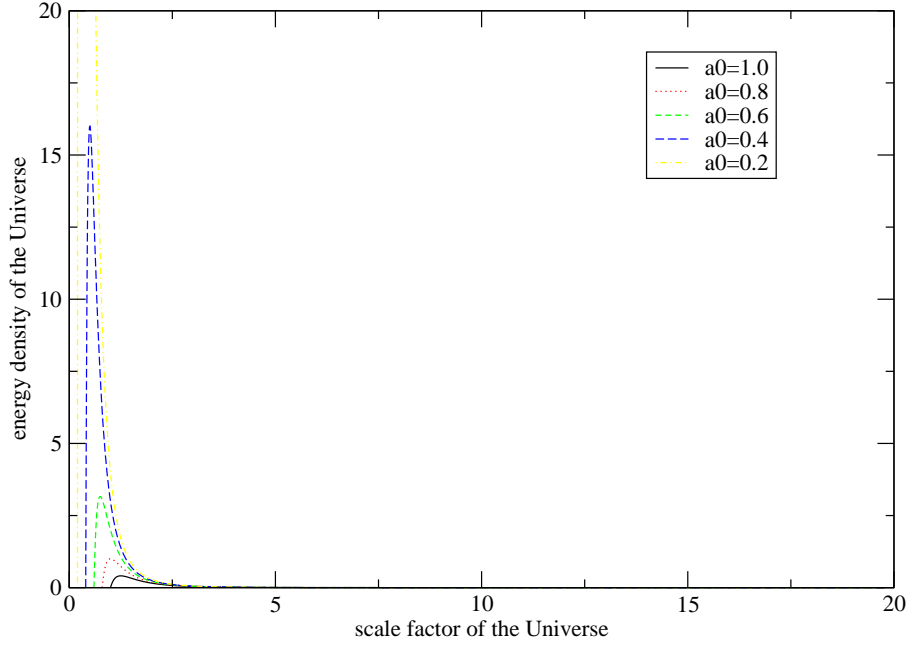}
\caption{The numerically calculated relation between $a$ and $\rho$ 
in $a\ge 0$ and $\bar{\rho}\ge 0$ region. Here $a0$ denotes the value of 
the scale factor at when the energy density of the Universe vanishes. 
In this figure $K=0$ and $w=0$ are used.}
\end{figure}
This differential equation can be solved numerically in $a\ge 0$ and 
$\bar{\rho}\ge 0$ region, as a result Fig.3 is given. This result 
implies that if the quantum effect of QG is considered the energy 
density of the early Universe is free from the initial singularity.
\item $t$ - $a$ relation\\
\begin{figure}
\includegraphics{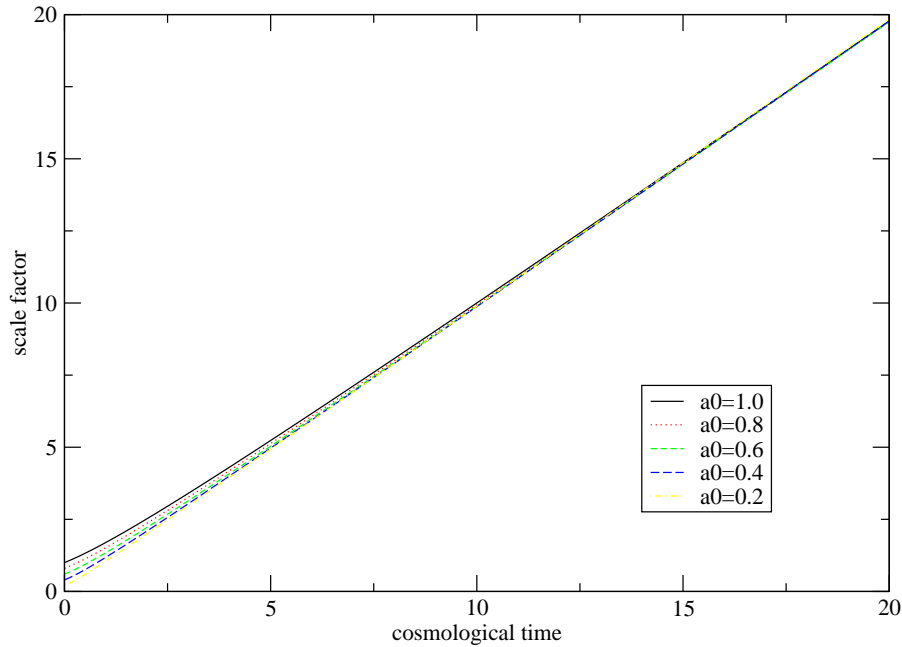}
\caption{The $t$ dependence of the scale factor $a$ in $t\ge 0$. Here 
$a0$ is the same as Fig.3, because of the uniquely determination of 
$t=0$. In this figure $K=0$ and $w=0$ are used, too.} 
\end{figure}
The cosmological time $t$ dependence of the scale factor $a$ is derived 
from Eqs.(\ref{link11}) and (\ref{link13}). Fig.4 gives the result. Here 
the requirement that $(da/dt)^2$ is kept positive in $\bar{\rho}\ge 0$ 
region with respect to the expansion of the Universe leads that the 
condition 
\begin{equation}
\frac{1}{3}-\frac{K}{\widetilde{\lambda}_*}\ge 0.
\end{equation}
If this is satisfied, the origin of the cosmological time $t=0$ can be 
determined uniquely at when the energy density of the Universe $\rho$ 
vanishes. Then the $t$ dependence of the scale factor in the very early 
Universe is considered as approximate linear because of Fig.5. This is 
different from the usual behavior of $a\propto\sqrt{t}$ in the 
radiation dominated period. As a result the horizon problem weakens, 
but to solve perfectly the other mechanism such as inflation is needed.
\begin{figure}
\includegraphics{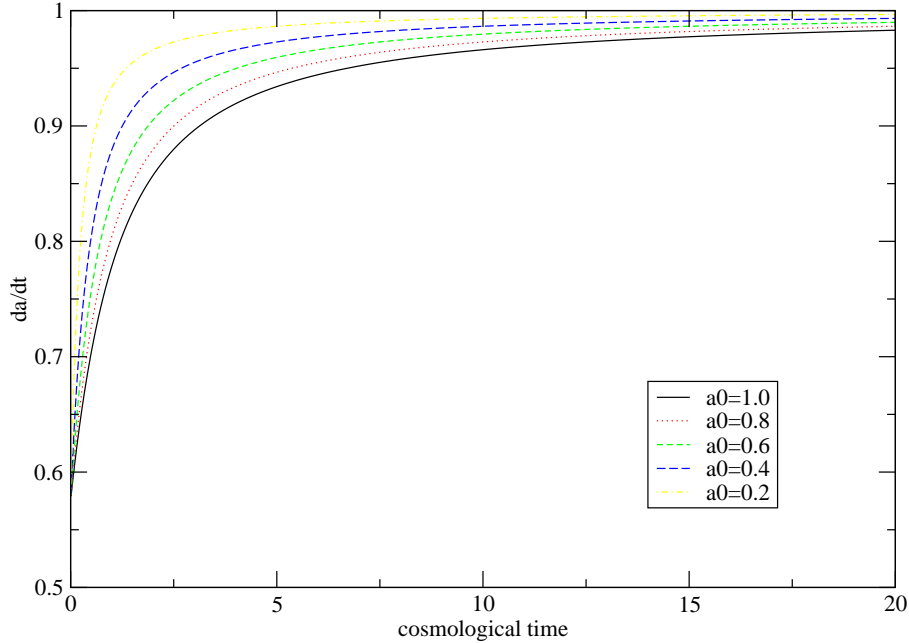}
\caption{The $t$ dependence of the gradient of the scale factor.}
\end{figure}
\end{enumerate}

\section{Conclusion}
In general to identify the energy scale is the most important task, 
if the result of ERGE analysis tries to apply the actual Universe.
The ansatz Eq.(\ref{link10}) used in this paper is considered to work 
effectively at several points. As a result the ERGE modified Universe 
has the advantage of the followings.
\begin{itemize}
\item The Universe is free from the singularities.
\item The origin of the cosmological time is uniquely decided.
\end{itemize}
In the next stage to solve the horizon and flatness problems and so on 
the other mechanics such as inflation is needed. Then this Universe 
is thought of as the credible framework in which the inflation 
mechanics are constructed and checked because of including the quantum 
effect explicitly. 

By the way the several assumptions are used implicitly. They are that to 
apply the Euclidean ERGE results to the Minkowski Universe, to apply 
the pure gravity results to the Universe coupled with matter. These 
assumptions are necessarily to be investigated in detail. 

At last the minimum scale factor of the Universe is natural decided. 
The ratio of this characteristic value to the current value of the 
scale factor is maybe related to Planck scale. Perhaps the origin 
of the quantum effect is also found in the very early Universe.


\begin{thebibliography}{99}
\bibitem{Deser}S.Deser, Annalen Phys. 9 (2000) 299-307
\bibitem{Kawai}H.Kawai,M.Ninomiya, Nucl. Phys. {\bf B336} (1990) 115
\bibitem{Reuter}M.Reuter, Phys. Rev. {\bf D57} (1998) 971
\bibitem{Souma}W.Souma, Prog.Theor.Phys. 102 (1999) 181-195
\bibitem{Weinberg}S.Weinberg,
{\it Ultraviolet divergences in quantum theories of gravitation}, ed. S.Hawking
and W.Israel(Cambridge Univ. Press,Cambridge 1979)
\bibitem{Percacci}R.Percacci,D.Perini, Phys. Rev. {\bf D68} (2003) 044018
\bibitem{Reuter2}M.Reuter,F.Saueressig, Phys. Rev. {\bf D65} (2002) 065016
\bibitem{Souma2}W.Souma, {\tt gr-qc/0006008}
\end{thebibliography}
\end{document}